\documentclass[twocolumn,showpacs,aps]{revtex4}
\usepackage{bm}

\begin{document}
\title{Consistent classical and quantum mixed dynamics}
\author{Michael J. W. Hall}
\affiliation{Theoretical Physics, IAS, \\ Australian National
University,\\
Canberra ACT 0200, Australia}

\begin{abstract}
A recent proposal for mixed dynamics of classical and quantum ensembles is shown, in contrast to other proposals, to satisfy the minimal algebraic requirements proposed by Salcedo for any consistent formulation of such dynamics. Generalised Ehrenfest relations for the expectation values of classical and quantum observables are also obtained.  It is further shown that additional desirable requirements, related to separability, may be satisfied under the assumption that only the configuration of the classical component is directly accessible to measurement, eg, via a classical pointer. Although the mixed dynamics is formulated in terms of ensembles on configuration space, thermodynamic mixtures of such ensembles may be defined which are equivalent to canonical phase space ensembles on the classical sector.  Hence, the formulation appears to be both consistent and physically complete.
\end{abstract}

\maketitle

\section{Introduction}

Many proposals have been made for mixing classical and quantum dynamics \cite{tries,royal}.  These proposals may be broadly classified into mean-field, phase-space and trajectory categories,  and all fail some important criterion - such as conservation of energy and probability, back-reaction by the quantum component on the classical component, positivity of probabilities, correct equations of motion in the limit of no interaction, and describing all interactions of physical interest \cite{royal,sud2,boucher, sal1,sal2, ternoperes, salcedocom,sahoo, sal3,agostini}).   Hence, while often of practical interest for calculations in molecular dynamics \cite{sal3,agostini,calc}, the above proposals have not led to a formulation of mixed classical and quantum dynamics that may be regarded as physically fundamental.  Such a formulation would not only be of interest for making physically-consistent numerical calculations, but also for modelling `classical' measurement apparatuses and coupling quantum systems to classical spacetime metrics \cite{boucher}.

Recently, a new proposal has been made that falls outside the above categories \cite{hr}. It is based on the description of physical systems by ensembles on configuration space, and satisfies all of the abovementioned criteria.  Moreover, it has been successfully applied to discuss position and spin measurements, interacting classical and quantum oscillators, and the coupling of quantum fields to classical spacetime \cite{hr,kiefer}.  However, while this `configuration-ensemble' approach is therefore a promising formulation of mixed dynamics, only particular observable properties, such as position, momentum, energy and angular momentum, have been discussed in any detail. Hence, further investigation is necessary to determine whether it is fully self-consistent.

In this regard, Salcedo has recently specified two ``{\it minimal requirements for a consistent classical-quantum formulation}'' \cite{sal3}:
\begin{description}
\item[(i)] a Lie bracket may be defined on the set of observables, and
\item[(ii)] the Lie bracket is equivalent to the classical Poisson bracket for any two classical observables, and to $(i\hbar)^{-1}$ times the quantum commutator for any two quantum observables.
\end{description}
These have been previously justified on physical grounds by Caro and Salcedo \cite{sal2}, and are higly nontrivial: none of the abovementioned proposals, in those cases where they have been sufficiently developed to identify the general `classical' and `quantum' observables of the theory, satisfy both requirements \cite{sal2,sal3,agostini}.  These minimal requirements therefore provide a critical test for the configuration-ensemble approach to mixed dynamics.

In Sec.~II it is shown that the configuration-ensemble formulation, unlike other proposals, {\it does} pass the above test. In particular, a Lie algebra of observables may be defined that satisfies requirements (i) and (ii) above.  

It is further shown in Sec.~II that the quantum Ehrenfest relations generalise to mixed systems, and in particular that the expectation values for the position and momentum observables of linearly-coupled classical and quantum oscillators obey the classical equations of motion. It follows that the configuration-ensemble formulation also satisfies a ``{\it definite benchmark ... for an acceptable classical-quantum hybrid formalism}'' proposed by Peres and Terno \cite{ternoperes}.

In Sec.~III it is demonstrated that the additional reasonable requirement
\begin{description}
\item[(iii)] The classical configuration is invariant under any canonical transformation applied to the quantum component, and vice versa
\end{description}
is also satisfied.  This  requirement is weaker than a related requirement, that the Lie bracket vanishes for {\it all} pairs of classical and quantum observables, previously proposed by Caro and Salcedo \cite{sal2} (see equation (18) thereof), but is sufficient if the physically reasonable assumption is made that only the configuration of the classical component is directly accessible by measurement (eg, via a classical pointer).

Finally, in Sec.~IV it is shown that the configuration-ensemble approach is consistent with thermodynamics.  In particular, a generalised `canonical ensemble' may be defined as a suitable mixture of distinguishable stationary ensembles on configuration space, which is equivalent to the usual canonical ensemble on phase space for any semi-ergodic classical system, and to the usual canonical ensemble on Hilbert space for any quantum system.  This demonstrates that the formulation of the approach on configuration space is not a barrier to describing all physically relevant systems (and also implies classical statistical mechanics may be given a Hamilton-Jacobi formulation).

\section{Observables in the configuration-ensemble approach}

\subsection{General observables}

The description of physical systems by ensembles on configuration space may be introduced at quite a fundamental and generic level \cite{hr,super}.  The starting point is simply a  probability density $P$ on the configuration space of the system, the dynamics of which is assumed to satisfy an action principle.  Thus, there is a canonically conjugate quantity, $S$, on the configuration space, and an `ensemble' Hamiltonian, $\tilde{H}[P,S]$, such that
\begin{equation} \label{motion}
\frac{\partial P}{\partial t} = \frac{\delta \tilde{H}}{\delta S},~~~~~\frac{\partial S}{\partial t} = - \frac{\delta \tilde{H}}{\delta P}.
\end{equation}
Here $\delta/\delta f$ denotes the appropriate variational derivative on the configuration space.  For the particular case of a continuous configuration space, indexed by `position coordinate' $\xi$, and a functional $L[f]$ of the form $L[f]=\int d\xi\,F(f,\nabla f,\xi)$, one has the useful formula  \cite{goldstein}
\[ \delta L/ \delta f = \partial F/\partial f - \nabla \cdot[\partial F/\partial (\nabla f)] . \]

As a simple example, the ensemble Hamiltonian
\[ \tilde{H}_Q = \int dq\, P\left[ \frac{|\nabla  S|^2}{2m}+ \frac{\hbar^2}{8m}\frac{|\nabla   P|^2}{P^2} +V(q) \right]  \]
describes a quantum spin-zero particle of mass $m$, moving under a potential $V(q)$.  In particular, the equations of motion (\ref{motion}) reduce in this case to the real and imaginary parts of the Schr\"{o}dinger equation
\[  i\hbar(\partial\psi/\partial t) =  \left[ -(\hbar^2/2m)\nabla^2 + V\right]\psi , \]
where $\psi(q,t):=P^{1/2}e^{iS/\hbar}$.  Moreover, if the limit $\hbar\rightarrow 0$ is taken in the ensemble Hamiltonian $\tilde{H}_Q$, the equations of motion for $P$ and $S$ reduce to the Hamilton-Jacobi and continuity equations for an ensemble of classical particles \cite{hr}.  Thus, in the configuration-ensemble approach, the primary difference between quantum and classical evolution lies in the choice of the ensemble Hamiltonian.  

It may be noted that the numerical value of the above ensemble Hamiltonian is just the quantum average energy, i.e.,
\[  \tilde{H}_Q[P,S] = \int dq\, \psi^*(q)\left[ -(\hbar^2/2m)\nabla^2 + V\right]\psi(q) , \]
as first demonstrated by Madelung in Eq.~(5') of Ref.~\cite{mad}.  The fact that the quantum equations of motion can be generated by regarding $P$ and $S$ as canonically conjugate fields appears to have first been noted by Bohm, following Eq.~(9) of Ref.~\cite{bohm1}.  Examples of mixed quantum-classical ensemble Hamiltonians have been given elsewhere \cite{hr,kiefer}, and also in the following sections.

While a number of general properties and applications of the configuration ensemble formalism have been previously considered \cite{hr}, the description of `observables' has only been briefly alluded to, with emphasis on particular quantities such as position, momentum, spin and energy.  The general description is therefore addressed here, to enable the properties of observables for mixed configuration ensembles to be compared against the two minimal requirements discussed in the Introduction.

Note first that the conjugate pair $(P,S)$ allows a Poisson bracket to be defined for any two functionals $A[P,S]$ and $B[P,S]$, via  \cite{goldstein}
\begin{equation} \label{poiss}  
\{A,B\} := \int d\xi\, \left( \frac{\delta A}{\delta P} \frac{\delta B}{\delta S} -\frac{\delta B}{\delta P} \frac{\delta A}{\delta S}  \right)  
\end{equation}
(where integration is replaced by summation over any discrete parts of the configuration space). Thus, the equations of motion (\ref{motion}) may be written as $\partial P/\partial t=\{P,\tilde{H}\}$ and $\partial S/\partial t=\{S,\tilde{H}\}$, and more generally it follows that
\begin{equation} \label{evo} 
d A/d t = \{A,\tilde{H}\} +\partial A/\partial t 
\end{equation}
for any functional $A[P,S,t]$.  The Poisson bracket is well known to be a Lie bracket \cite{goldstein}, and in particular is linear, antisymmetric, and satisfies the Jacobi identity.  Hence, the first minimal consistency requirement, given in Sec.~I above, is automatically satisfied by choosing the observables to be any set of functionals of $P$ and $S$ that is closed with respect to the Poisson bracket.  

It is important to note that an {\it arbitrary} functional $A[P,S]$ will not be allowable as an observable in general. This is analogous to the restriction of expectation values to bilinear forms of linear Hermitian operators in standard quantum mechanics, even though a commutator bracket can be more generally defined \cite{weinberg}. For example, the infinitesimal canonical transformation 
\[  P \rightarrow P + \epsilon \,\delta A/\delta S,~~~~~S \rightarrow S - \epsilon\, \delta A/\delta P \]
generated by any observable $A$  must preserve the interpretation of $P$ as a probability density, i.e., the normalisation and positivity of $P$ must be preserved.  This imposes the respective fundamental conditions \cite{hr}
\begin{equation} \label{con1} 
A[P,S+c] = A[P,S],~~~~~\delta A/\delta S = 0 {\rm ~if~} P(\xi)=0,   
\end{equation}
where $c$ is an arbitrary constant.  

Note that each of the above conditions is consistent with the Poisson bracket. First, defining $I[P,S]:=\int d\xi\, P$, the normalisation condition is simply the requirement that $I$ is invariant under allowed canonical transformations, i.e., that $\delta I= \epsilon\{I,A\}=0$.  Hence, if it holds for two observables $A$ and $B$, then it automatically holds for $\{ A,B\}$ via the Jacobi identity, since
\[ \{I,\{A,B\}\} = - \{A, \{B,I\}\}- \{B,\{I,A\}\}  = 0.  \]
Similarly, the positivity condition may be rewritten as $\delta P = \epsilon\{P,A\}=0$ whenever $P(\xi)=0$ (otherwise $P(\xi)$ can be decreased below 0 by choosing the sign of $\epsilon$ appropriately), which again  holds for $\{A,B\}$, if it holds for $A$ and $B$, as a consequence of the Jacobi identity.  

It follows that the observables corresponding to a configuration-ensemble description should be chosen as some set of functionals satisfying the normalisation and positivity constraints (\ref{con1}), that is closed with respect to the Poisson bracket in Eq.~(\ref{poiss}).  Under any such choice, the first minimal requirement~(i) in Sec.~I is automatically satisfied, where the Lie bracket is identified with the Poisson bracket.

\subsection{Classical and quantum observables}

 To determine whether the second minimal requirement~(ii) in Sec.~I is also satisfied, it is necessary to first define the `classical' and `quantum' observables of a mixed quantum-classical ensemble.
 
Consider, therefore, a mixed quantum-classical ensemble, indexed by the joint configuration $\xi=(q,x)$, where $q$ labels the quantum configuration and $x$ labels the classical configuration.  For example, $q$ may refer to the position of a quantum system, or, more generally, label some complete set of kets $\{|q\rangle\}$ \cite{super}.  In contrast, $x$ will always be taken here to refer to some continuous set of coordinates on a classical configuration space (to enable the discussion of classical phase space relations).  Hence, the mixed ensemble is described by two conjugate quantities $P(q,x,t)$ and $S(q,x,t)$.  

First, for any real classical phase space function $f(x,k)$, define the corresponding classical observable $C_f$ by
\begin{equation} \label{cf}  
C_f := \int dq\,dx\, P \, f(x,\nabla_x S)   
\end{equation}
(where integration with respect to $q$ is replaced by summation over any discrete portions of the quantum configuration space).  This is similar in form to a classical average, and hence the  numerical value of $C_f$ will be identified with the predicted expectation value of the corresponding function $f(x,k)$, i.e., \[ C_f \equiv \langle f\rangle. \]
It is easily checked that $C_f$ satisfies the required normalisation and positivity conditions (\ref{con1}).  Note that, for classical observables, $\nabla_x S$ plays the role of a momentum associated with the configuration $(q,x)$.

The Poisson bracket of any two classical observables $C_f$ and $C_g$ follows, using Eq.~(\ref{poiss}) and integration by parts with respect to $x$, as
\begin{eqnarray} \nonumber
\{C_f,C_g\} &=& \int dq\, dx\,\left[ -f\nabla_x\cdot(P\nabla_kg) + g\nabla_x\cdot (P\nabla_k f)\right]\\ \nonumber
&=& \int dq\, dx\,P\left( \nabla_x f\cdot\nabla_k g - \nabla_x g\cdot\nabla_k f \right)\\
\label{cbrack}
&=& C_{\{f,g\}} ,
\end{eqnarray}
where all quantities in the integrands are evaluated at $k=\nabla_x S$, and $\{f,g\}$ denotes the usual Poisson bracket for phase space functions.  Hence, the Lie bracket for classical observables is equivalent to the usual phase space Poisson bracket, as required.  Given that the observables are evaluated on a configuration space, rather than on a phase space, this is a somewhat remarkable result.

Second, for any Hermitian operator $M$ acting on the Hilbert space spanned by the kets $\{|q\rangle\}$, define the corresponding quantum observable $Q_M$ by
\begin{eqnarray} \label{qm}
Q_M &:=& \int dq\,dx\, \psi^*(q,x) M\psi(q,x) \\ \nonumber
 &=& \int dq\,dq'\,dx\,(PP')^{1/2}e^{i(S-S')/\hbar} \langle q'|M|q\rangle , 
 \end{eqnarray}
where $\psi(q,x):=P(q,x)^{1/2}e^{iS(q,x)/\hbar}$, $P=P(x,q)$, $P'=P(x,q')$, etc (and where integration with respect to $q$ and $q'$ is replaced by summation over any discrete portions of the quantum configuration space).  This is similar in form to a quantum average, with respect to the `hybrid wavefunction' $\psi(q,x)$, and hence the numerical value of $Q_M$ will be identified with the predicted expectation value of the corresponding operator $M$, i.e., 
\[ Q_M \equiv \langle M\rangle . \] 
It follows immediately from the second equality in Eq.~(\ref{qm}) that $Q_M$ satisfies the  normalisation and positivity conditions (\ref{con1}).

To evaluate the Poisson bracket of any two quantum observables $Q_M$ and $Q_N$, it is convenient to first express the Poisson bracket in terms of the hybrid wavefunction $\psi(q,x)$ and its complex conjugate $\psi^*(q,x)$.  One has in particular for any real functional $A[P,S]$ that
\[ \frac{\delta A}{\delta P} = \frac{\partial\psi}{\partial P} \frac{\delta A}{\delta \psi}+\frac{\partial\psi^*}{\partial P} \frac{\delta A}{\delta \psi^*} = \frac{1}{\psi^*\psi}\,{\rm Re}\left\{\psi\frac{\delta A}{\delta \psi}\right\} ,\]
\[ \frac{\delta A}{\delta S} = \frac{\partial\psi}{\partial S} \frac{\delta A}{\delta \psi}+ \frac{\partial\psi^*}{\partial S} \frac{\delta A}{\delta \psi^*} = -\frac{2}{\hbar}\,{\rm Im}\left\{\psi\frac{\delta A}{\delta \psi}\right\} ,\]
and hence, noting $-ad+bc={\rm Im}\{(a+ib)(c-id)\}$, that
\begin{equation} \label{psibrack}
\{ A,B\} = \frac{2}{\hbar}\, {\rm Im}\left\{\int dq\,dx\,\frac{\delta A}{\delta\psi}\, \frac{\delta B}{\delta\psi^*} \right\}  .
\end{equation}
Recalling that $M$ and $N$ are Hermitian, so that $\psi^* M\psi$ may be replaced by $(M\psi)^*\,\psi$ in equation (\ref{qm}), it immediately follows that
\begin{equation} \label{qbrack}
\{Q_M,Q_N\} = \frac{2}{\hbar}\, {\rm Im}\left\{\int dq\,dx\, (M\psi)^*N\psi \right\} = Q_{[M,N]/(i\hbar)} ,
\end{equation}
where $[M,N]$ denotes the usual quantum commutator $MN-NM$.  Hence, the Lie bracket for quantum observables is equivalent to the usual quantum commutator, as required.

Eqs.~(\ref{poiss}), (\ref{cbrack}) and (\ref{qbrack}) are the main results of this section.  They imply that any set of observables containing the quantum observables $C_f$ and the quantum observables $Q_M$, that is closed under the Poisson bracket, will satisfy both of the minimal requirements (i) and (ii) in Sec.~I for a consistent mixed quantum-classical formulation.  In the following sections, it will be shown that further desirable properties can also be accommodated within the configuration-ensemble formulation.

\subsection{Example: generalised Ehrenfest relations}

Consider a mixed quantum-classical ensemble, corresponding to  a quantum particle of mass $m$ interacting with a classical particle of mass $M$ via a potential $V(q,x)$, where $q$ and $x$ denote the position configurations of the quantum and classical particles respectively.  For simplicity, it will be assumed that both $q$ and $x$ are one-dimensional.  The mixed ensemble is therefore described by a probability density $P(q,x)$, a canonically conjugate field $S(q,x)$, and an ensemble Hamiltonian of the form \cite{hr}
\begin{eqnarray} \nonumber 
\tilde{H}_{QC}[P,S] &:=& \int dq\,dx\, P\left[ \frac{(\partial_q  S)^2}{2m}+ \frac{\hbar^2}{8m}\frac{(\partial_q   P)^2}{P^2}\right. \\ \label{hqc} 
&~&~~~~~ \left. + \frac{(\partial_x  S)^2}{2M} +V(q,x) \right] ,
\end{eqnarray}
where $\partial_q$ and $\partial_x$ denote the partial derivatives with respect to $q$ and $x$ respectively.

Now, the expectation values of the classical position and momentum observables follow from Eq.~(\ref{cf}) as
\[ \langle x \rangle = C_x = \int dq\,dx\, P\,x,~~~~\langle k\rangle = C_k = \int dq\,dx\, P\,\partial_xS ,\]
and the expectation values of the quantum position and momentum observables follow from Eq.~(\ref{qm}) as
\[ \langle q \rangle = Q_q = \int dq\,dx\, P\,q,~~~~\langle p\rangle = Q_p = \int dq\,dx\, P\,\partial_qS .\]

The evolution of these expectation values may be calculated via Eqs.~(\ref{evo}) and (\ref{hqc}), and, as shown below, one finds
\begin{equation} \label{ex}
\frac{d}{dt} \langle x\rangle = M^{-1}\langle k \rangle,~~~~\frac{d}{dt} \langle k\rangle = -\langle \partial_x V \rangle,
\end{equation}
\begin{equation} \label{eq}
\frac{d}{dt} \langle q\rangle = m^{-1}\langle p \rangle,~~~~\frac{d}{dt} \langle p\rangle = -\langle \partial_q V \rangle .
\end{equation}
These results are a clear generalisation of the standard Ehrenfest relations for quantum systems, and similarly imply that the centroid of a narrow initial probability density $P(q,x)$ will evolve classically, for short timescales at least.

Note that for the case of  linearly-coupled classical and quantum oscillators, with
\[ V(q,x) = \frac{1}{2}m\omega^2q^2 + \frac{1}{2}M\Omega^2 x^2 + Kqx, \]
Eqs.~(\ref{ex}) and (\ref{eq}) simplify to give the closed set of equations
\[ \frac{d}{dt} \langle x\rangle = M^{-1}\langle k \rangle,~~~~\frac{d}{dt} \langle k\rangle = -M\Omega^2 \langle x \rangle -K\langle q \rangle,\]
\[ \frac{d}{dt} \langle q\rangle = m^{-1}\langle p \rangle,~~~~\frac{d}{dt} \langle p\rangle = -m\omega^2 \langle q \rangle -K\langle x \rangle . \]
Thus, the centroid of the probability density obeys precisely the same equations of motion as two fully classical oscillators (or two fully quantum oscillators).  As noted in the Introduction, this `correspondence principle' has been previously proposed by Peres and Terno as a benchmark for any acceptable classical-quantum hybrid formalism \cite{ternoperes}.  The above result therefore further supports the consistency of the configuration-ensemble formulation of mixed dynamics.

To demonstrate the generalised Ehrenfest relations (\ref{ex}) and (\ref{eq}), note first that $\langle x\rangle$ has no explicit dependence on $S$ or $t$, and hence from Eqs.~(\ref{poiss}), (\ref{evo}) and (\ref{hqc}) one has
\begin{eqnarray*} 
\frac{d}{dt} \langle x\rangle &=& \int dq\,dx\, x\frac{\delta\tilde{H}_{QC}}{\delta S}\\
 &=&  -\int dq\,dx\, x \left[ m^{-1}\partial_q(P\partial_qS) + M^{-1}\partial_x(P\partial_xS) \right].
 \end{eqnarray*}
Applying integration by parts with respect to $q$ and $x$ to the first and second terms, respectively, yields the first relation in Eq.~(\ref{ex}).  The first relation in Eq.~(\ref{eq}) is obtained in a similar manner.

To obtain the remaining relations, note that
\begin{eqnarray*}
\frac{\delta \tilde{H}_{QC}}{\delta P} &=& \frac{(\partial_q  S)^2}{2m} + \frac{(\partial_x  S)^2}{2M} +V\\
&~& ~~+ \frac{\hbar^2}{2m}\frac{\partial P^{1/2}}{\partial P} \frac{\delta}{\delta P^{1/2}} \int dq\,dx\,  (\partial_q P^{1/2})^2 .
\end{eqnarray*}
The final term simplifies to
$-(\hbar^2/2m) (\partial_q^2 P^{1/2})/P^{1/2}$,
which may be recognised as the so-called `quantum potential' \cite{bohm1}, and
Eqs.~(\ref{poiss}), (\ref{evo}) and (\ref{hqc}) then yield
\begin{eqnarray*}
\frac{d}{dt} \langle k\rangle &=& -\int dq\,dx\, (\partial_x S) \left[ \partial_q\left(P\frac{\partial_qS}{m}\right) + \partial_x\left(\frac{P\partial_xS}{M}\right) \right]\\
&~&+ \int dq\,dx\, (\partial_x P)\left[ \frac{(\partial_q  S)^2}{2m} + \frac{(\partial_x  S)^2}{2M} +V \right]\\
&~& -(\hbar^2/2m)\int dq\,dx\, (\partial_x P)(\partial_q^2 P^{1/2})/P^{1/2} .
\end{eqnarray*}
Applying integration by parts to the first line, with respect to $q$ and $x$ for the first and second terms thereof respectively, and applying integration by parts to the second line with respect to $x$, leads to cancellation of all terms involving $S$, with a term $-\int dqdx\,P\partial_xV$ remaining.  Since the integral in the third line is proportional to
\[ \int dq\,dx\, (\partial_x P^{1/2})(\partial_q^2 P^{1/2}) = -\frac{1}{2}\int dq\,dx\,\partial_x\left(\partial_qP^{1/2}\right)^2  \]
which vanishes identically, the second relation in Eq.~(\ref{ex}) immediately follows.  Finally, the second relation in Eq.~(\ref{eq}) is obtained by similar reasoning (replacing $\partial_q$ by $\partial_x$ in  appropriate places).

\section{Separability and measurement}

In contrast to other proposals for mixed dynamics, the configuration-ensemble approach has been shown to pass the critical test of meeting the two minimal consistency requirements (i) and (ii) in Sec.~I.  However, while the approach thereby gains a special status, there are a number of other requirements that may reasonably be expected of a physical theory.  Some of these, related to local aspects of separability and measurement, are discussed below.

\subsection{Configuration separability}

It is natural to expect that the classical configuration is invariant under any canonical transformations applied solely to the quantum system, and vice versa.  This means, in particular, that a measurement of the classical configuration cannot detect whether or not a transformation has been applied to the quantum system, and vice versa, when the components are noninteracting.  This property of `configuration separability' corresponds to requirement (iii) of the Introduction.  

To show that this property indeed holds, consider first a canonical transformation generated by an arbitrary quantum observable $Q_M$.  Any classical observable depending only on the classical configuration $x$ is of the form $C_g$ with $g=g(x)$ (eg, $C_x$).  One thus has $C_g=\int dq\,dx\,Pg(x)=\int dq\,dx\,\psi^*\psi\,g(x)$, and it follows immediately via Eq.~(\ref{psibrack}) that
\begin{equation} \label{cx}
\{ C_{g(x)},Q_M\} = \frac{2}{\hbar} \,{\rm Im}\left\{ \int dq\,dx\,\psi^*g(x)M\psi\right\} = 0 ,
\end{equation}
since $M$ acts only on the quantum component of the hybrid wavefunction and so commutes with $g(x)$.  Hence, the expectation value of $C_g$ is not changed by the transformation.  Similarly, consider a canonical transformation of the classical component, generated by an arbitrary classical observable $C_f$.  Any quantum observable depending only on the quantum configuration $q$ is of the form $Q_G$ with $G=G({q})$.  Thus, $Q_G=\int dq\,dx\, \psi^*G({q})\psi=\int dq\,dx\,P\,G(q)$, and it follows immediately via Eq.~(\ref{poiss}) that
\begin{equation} \label{qq}
\{ Q_{G({q})},C_f \} = - \int dx\,dq\, G(q)\nabla_x\cdot \left(P\nabla_k f\right) = 0,
\end{equation}
using integration by parts with respect to $x$.

Eqs~(\ref{cx}) and (\ref{qq}) show that configuration separability is satisfied.  Further, if $q_m$ and $p_m$ label the $m$th position and momentum coordinates of a quantum particle, and $x_m$ and $k_m$ similarly label the $m$th position and momentum coordinates of a classical particle, then, noting that $Q_{p_m}=\int dq\,dx\, P(\partial S/\partial q_m)$ and using Eq.~(\ref{poiss}), 
\[ \{ C_{k_m},Q_{p_n}\} = \int dq\,dx\,\left( -\frac{\partial S}{\partial x_m} \frac{\partial P}{\partial q_n} + \frac{\partial S}{\partial q_n} \frac{\partial P}{\partial x_m} \right) = 0, \]
using integration by parts with respect to $q$ and $x$ on the first and second terms, respectively (it can also be shown that $\{ C_{k_m},Q_M\}=\{ C_f,Q_{p_m}\}=0$).  Together with Eqs.~(\ref{cx}) and (\ref{qq}), this yields the properties
\begin{equation}  \{ C_{x_m}, Q_{q_n} \} = \{ C_{x_m}, Q_{p_n}\} = 0, 
\end{equation}
\begin{equation} 
\{ C_{k_m},Q_{q_n}\} = \{ C_{k_m},Q_{p_n}\} =0 , 
\end{equation}
for all $m$ and $n$.  Noting Eqs.~(\ref{cbrack}) and (\ref{qbrack}), these properties correspond to  conditions assumed by Salcedo \cite{sal1} (in Eq.~(1) thereof) for proving a no-go theorem for mixed classical and quantum dynamics.  However, this theorem is inapplicable here, as a further required condition, that observables are generated by a product algebra, does not hold (see also Sec.~V below).

\subsection{Strong separability}

It is important to note that while configuration separability holds, as per Eqs~(\ref{cx}) and (\ref{qq}), the stronger separability requirement 
\begin{equation} \label{strong}
\{ C_f, Q_M\} = 0 ~~~?
\end{equation}
does {\it not} hold for arbitrary classical and quantum observables.  For example, for classical and quantum particles having masses $m$ and $m'$ respectively, the Poisson bracket of the `kinetic energy' observables corresponding to $f(x,k)=|k|^2/(2m)$ and $M=-\hbar^2|\nabla_q|^2/(2m')$ is given by
\[ \{C_f,Q_M \} = \frac{\hbar^2}{2mm'}\int dq\,dx\,P(\nabla_x S)\cdot \nabla_x (P^{-1/2}\nabla_q^2P^{1/2}) , \]
which does not vanish identically for arbitrary $P$ and $S$.  However, it will be argued here that the violation of `strong separability' does not necessarily lead to any physical inconsistencies.

First, it should be noted that in the particular case where the mixed system describes quantum matter coupled to classical spacetime \cite{boucher,hr}, a failure of Eq.~(\ref{strong}) is irrelevant to separability issues, as there is no sense in which interaction between the systems can be `switched off' - matter bends space and space curves matter, and so a change in one component is fully expected to drive a change in the other component.  This corresponds to the direct coupling of the metric tensor to the fields in the corresponding ensemble Hamiltonian \cite{hr}, and there is no sense in which this ensemble Hamiltonian can be reduced to a simple sum of a classical and a quantum contribution.  

Second, it is important to note that Eq.~(\ref{strong}) does hold in the special case that the classical and quantum components are {\it independent} \cite{hr}, i.e.,  when
\begin{equation} \label{fac}
P(q,x) = P_Q(q) \,P_C(x),~~~S(q,x) = S_Q(q) + S_C(x).
\end{equation}
Thus, independent ensembles are fully described by two conjugate pairs $(P_Q,S_Q)$ and $(P_C,S_C)$ corresponding to the quantum and classical components respectively.  
To demonstrate Eq.~(\ref{strong}) for this case, note first from Eq.~(\ref{cf}) that 
\[ C_f=\int dq\,dx\,\psi^*\psi\, f(x,k),~~~k = \frac{\hbar}{2i}\left( \frac{\nabla_x \psi}{\psi} - \frac{\nabla_x \psi^*}{\psi^*} \right) , \]
and hence that
\begin{eqnarray*} 
\frac{\delta C_f}{\delta \psi} &=& \psi^*f+\psi^*\psi\,(\nabla_k f)\cdot \frac{\partial k}{\partial\psi}\\
&~~& - \nabla_x\cdot\left(\psi^*\psi\,(\nabla_k f)\cdot\frac{\partial k}{\partial(\nabla_x\psi)}\right)\\
&=& \psi^*f-\frac{\hbar}{2i}\frac{\psi^*}{\psi}\left(\nabla_k f\cdot \nabla_x\psi \right) -\frac{\hbar}{2i}\nabla_x \cdot \left(\psi^*\nabla_k f\right) .
\end{eqnarray*}
Moreover, from Eq.~(\ref{qm}) one has $\delta Q_M/\delta\psi^* = M\psi$.  Since Eq.~(\ref{fac}) is equivalent to a factorisation $\psi(q,x)=\psi_Q(q)\psi_C(x)$ of the hybrid wavefunction, implying that $k\nabla_x S_C$ is independent of $q$, it follows that 
\[ \int dq\,dx\, \frac{\delta C_f}{\delta \psi}\frac{\delta Q_M}{\delta\psi^*} = \int dq\,\psi^*_QM\psi_Q \left\{ \int dx\,\psi^*_C\psi_Cf \right. ~~~~~~~~\] 
\[ ~~~~~~~~~~~ - \left. \frac{\hbar}{2i}\int dx\,\left[\psi^*_C(\nabla_kf\cdot\nabla_x\psi_C) +  \psi_C\nabla_x\cdot (\psi^*_C\nabla_kf)\right] \right\} \]
\[ ~~~~~~~~~~~~~~= \int dq\,\psi_Q^*M\psi_Q \int dx\,\psi^*_C\psi_Cf ,\]
where integration by parts has been used to obtain the final result.  This expression is clearly real, implying immediately from Eq.~(\ref{psibrack}) that $\{C_f,Q_M\} =0$, as required.  

Thus, strong separability is satisfied for independent ensembles.  Further, since such ensembles remain independent under ensemble Hamiltonians of the form $\tilde{H}= C+Q$ (for some pair of classical and quantum observables $C$ and $Q$), strong separability holds at all times for noninteracting independent ensembles.

More generally, the violation of strong separability poses an apparent problem:  a transformation acting solely on one component can lead to changes in the expectation values of observables in the other component.  However, this problem may be resolved by imposing a physically reasonable restriction on the type of observables which are directly accessible to measurement.

For example, consider the assumption:
\begin{description}
\item[(A)] The only observables accessible to direct measurement are classical configuration observables.
\end{description}
Here `configuration observables' are those which depend only on the configuration of the classical system, i.e., of the form $C_{g(x)}$.  Under this assumption, information about any other classical or quantum observables is obtainable only indirectly, by coupling them to such a classical configuration observable.  
 
 Empirically, this assumption is a very reasonable one to make, as in practice all measurements do reduce to the observation of some position or configuration of a classical apparatus (such as the position of a pointer).  It is of interest to note that a similar assumption (for different reasons) is also made in the deBroglie-Bohm interpretation of standard quantum mechanics \cite{bohm1,bohm}.
 
Now, under assumption (A), only changes in the expectation values of classical configuration observables are directly observable.  However, from Eq.~(\ref{cx}) above, such expectation values are invariant under any canonical transformations that act solely on the quantum component.  Hence, the assumption implies that violations of strong separability are simply not observable.

\subsection{Measurement aspects}

Examples of measurements of position and spin on quantum ensembles, via interaction with an ensemble of classical pointers, and the corresponding decoherence of the quantum component relative to the classical component, have been described previously within the configuration-ensemble approach \cite{hr}. Here it is noted there is a simple model for describing the indirect measurement of any quantum observable, via a direct measurement of the configuration of a classical measuring apparatus. The existence of such a model indicates that assumption (A), discussed in Sec.~III~B above, does not restrict the types of information that can be gained by measurement.

In particular, the measurement of an arbitrary quantum observable, $Q_M$, may be modelled by the ensemble Hamiltonian
\[ \tilde{H} := \tilde{H}_0+\kappa(t) \int dq\,dx\, \psi^*(q,x) \left(\frac{\hbar}{i}\frac{\partial}{\partial x}\right) M\psi(q,x)  ,  \]
where $\kappa(t)$ vanishes outside the measurement period, and $x$ denotes the position of a one-dimensional classical pointer (with integration over $q$ replaced by summation over any discrete values).  Note that the interaction term satisfies the normalisation and positivity constraints (\ref{con1}).  As will be shown, this term correlates the eigenvalues of $M$ with the position of the pointer, in a manner rather similar to purely quantum models of measurement.  

It is convenient to assume that the measurement takes place over a sufficiently short time period, $[0,T]$, such that $\tilde{H}_0$ can be ignored during the measurement. The equations of motion during the interaction then follow via Eqs.~(\ref{motion}) and (\ref{psibrack}) as being equivalent to the hybrid Schr\"{o}dinger equation
\[ i\hbar\frac{\partial\psi}{\partial t} = \kappa(t) \left(\frac{\hbar}{i}\frac{\partial}{\partial x}\right)M\psi . \]
For an initially independent ensemble at time $t=0$, as per Eq.~(\ref{fac}), this equation may be trivially integrated to give 
\begin{equation} \label{sum} 
\psi(q,x,T) = \sum_n c_n\, \psi_C(x-K\lambda_n)\, \langle q|n\rangle  
\end{equation}
at the end of the measurement interaction, where $K=\int_0^T dt\,\kappa(t)$, $|n\rangle $ denotes the eigenstate corresponding to eigenvalue $\lambda_n$ of $M$, and $c_n = \langle n|\psi_Q\rangle$ (with $|\psi_Q\rangle$ defined via $\langle q|\psi_Q\rangle:=\psi_Q(q)$).  It has been assumed for simplicity here that $M$ is nondegenerate (the degenerate case is considered further below). 

The pointer probability distribution after measurement follows immediately from Eq.~(\ref{sum}) as
\begin{equation} \label{point}  
P(x,T) = \int dq\,\psi^*\psi = \sum_n |c_n|^2 P_C(x-K\lambda_n)  . 
\end{equation}
Hence, the initial pointer distribution is displaced by an amount $K\lambda_n$ with probability $|c_n|^2$, thus correlating  the position of the pointer with the eigenvalues of $M$.  In particular, choosing a sufficiently narrow initial distribution $P_C(x)$ (eg, a delta-function), the displaced distributions will be nonoverlapping (corresponding to a `good' measurement), and eigenvalue $\lambda_n$ will be perfectly correlated with the measured pointer position.  

The above shows that the indirect measurement of any quantum observable may be modelled via interaction with a strictly classical measuring apparatus, followed by a direct measurement of the classical configuration.  Note that `collapse' of the quantum component of the ensemble can also be modelled, {\it if} desired.  Suppose in particular that the `real' position of the pointer is determined to be $x=a$.  This must correspond to just one of the nonoverlapping distributions $P_C(x-K\lambda_n)$, and updating $P(q,x,T)$ via Bayes theorem implies, via Eqs.~(\ref{sum}) and (\ref{point}), that 
\[  P_a(q,x,T) = \delta(x-a) P(q,a,T)/P(a,T) = \delta(x-a)\,|\langle q|n\rangle|^2  .\]
Moreover, it is natural to update the conjugate quantity $S(q,x,t)$ via the minimal substitution
\[  S_a(q,x,t) = S(q,a,T) .  \]
Note that the `collapsed' ensemble after measurement is thus independent as per Eq.~(\ref{fac}) (with quantum component described by $\psi_a(q)=\langle q|n\rangle$ up to a phase factor).  Hence, strong separability as per Eq.~(\ref{strong}) is satisfied if the pointer and the quantum system do not interact after the measurement.

Finally, for a degenerate operator $M$ with eigenvalue decomposition $\sum_n \lambda_nE_n$, similar results are obtained, but with $c_n\langle q|n\rangle$ in Eq.~(\ref{sum}) replaced by $\langle q| E_n|\psi_Q\rangle$, $|c_n|^2$
by $p_n=\langle\psi_Q|E_n|\psi_Q\rangle$, and the `collapsed' quantum component by $\psi_a(q)=(p_n)^{-1/2}\langle q|E_n|\psi_Q\rangle$.

\section{Mixtures and thermodynamics}

It has been demonstrated above that the configuration-ensemble approach provides a consistent formulation of mixed quantum and classical dynamics.  However, given that this approach describes classical ensembles via a configuration space, rather than a phase space, this raises a potential completeness issue:  do all classical phase space ensembles have a counterpart in this approach?  Further, given the lack of a natural phase space entropy, can the configuration-ensemble approach deal with thermal ensembles in a manner that is compatible with both classical and quantum thermodynamics?  It is shown briefly below that both these questions have positive answers.

The central concept required is that of a {\it mixture} of configuration ensembles.  In particular, if a physical system is described by the configuration ensemble $(P_j,S_j)$ with prior probability $p_j$, then it may be said to correspond to the  mixture $\{(P_j,S_j);p_j\}$.  For quantum ensembles, such mixtures are conveniently represented by density operators.  More generally, however, there is no similarly convenient representation.  The average of any observable $A[P,S]$ over a mixture is given by
\begin{equation} \label{ava} 
\langle A\rangle = \sum_j p_j\,A[P_j,S_j]  . 
\end{equation}

The first question posed above can now easily be answered, using the fact that any classical phase space point, $\gamma=(x',k')$, may be described by a classical configuration ensemble $(P_\gamma,S_\gamma)$, defined by
\[ P_\gamma(x):= \delta(x-x'), ~~~~~S_\gamma(x):= k'\cdot x  .  \]
In particular, the value of any classical observable $C_f$, corresponding to the average value of $f$, follows via Eq.~(\ref{cf}) as
\begin{equation} 
 C_f[P_\gamma,S_\gamma] = \langle f\rangle_\gamma = f(x',k')  .  
 \end{equation}
It follows immediately that {\it any} classical phase space ensemble, represented by some phase space density $p(x',k')$, may equivalently be described by the mixture $\{(P_\gamma,S_\gamma);p\}$ of classical configuration ensembles.

To address the second question above, one further requires the notions of `stationary' and `distinguishable' configuration ensembles.  First, stationary ensembles are those for which all observable quantities are time-independent,  and are characterised by the property \cite{hr}
\begin{equation} \label{stat} 
\partial P/\partial t = 0, ~~~~~\partial S/\partial t = -E, 
\end{equation}
for some constant $E$.  Second, two configuration ensembles are defined to be distinguishable if there is some observable which can distinguish unambiguously between them, i.e., the ranges of the observable for each ensemble do not overlap.  Thus, for example, two quantum ensembles are distinguishable if the corresponding wavefunctions are orthogonal, while two classical ensembles are distinguishable if the ranges of $(x,k)$ over the supports of the ensembles are nonoverlapping (with $k=\nabla_x S)$.  

A {\it thermal mixture} may now be defined as a mixture of distinguishable stationary ensembles $\{ (P,S);p(P,S|\tilde{H}) \}$ such that 
\begin{equation} \label{mix}
p(P,S|\tilde{H}) \sim e^{-\beta \tilde{H}[P,S]},~~~~~ \beta >0   . 
\end{equation}
This definition is, for present purposes, justified by its consequences, but it may also be motivated by appealing to properties of two distinct noninteracting systems in thermal equilibrium, described by joint ensemble Hamiltonian $\tilde{H}_T$, for which one expects 
\begin{equation} \label{two} p(PP',S+S'|\tilde{H}_T) = p(P,S|\tilde{H})\,p(P',S'|\tilde{H}')  ,
\end{equation}
for pairs $(P,S)$, $(P',S')$ of stationary ensembles.

For a quantum system with Hamiltonian operator $H$, the above definition immediately leads to the usual quantum canonical ensemble represented by the density operator proportional to $e^{-\beta H}$.  It will be shown below that, for an ergodic classical system with phase space Hamiltonian $H(x,k)$, the corresponding thermal mixture of configuration ensembles is equivalent to the classical canonical ensemble with phase space density proportional to $e^{-\beta H}$.  Note this result generalises immediately, via the factorisability of thermal mixtures in Eq.~(\ref{two}), to all `semi-ergodic' classical systems, i.e., to any classical system comprising noninteracting ergodic systems.  Thus, for example, since a one-dimensional oscillator is ergodic for any energy $E$, and higher dimensional oscillators can be decomposed into noninteracting normal modes \cite{goldstein}, any classical oscillator is a semi-ergodic system.

In particular, the classical ensemble Hamiltonian is 
\[  \tilde{H}[P,S] = C_H = \int dx\, P\, H[x,\nabla S] , \]
and it follows from  Eqs.~(\ref{motion}) and (\ref{stat}) that the stationary ensembles are then given by the solutions of the continuity and Hamilton-Jacobi equations 
\[  \nabla_x\cdot \left[ P\nabla_k H(x,\nabla_xS)\right]= 0,~~~   H(x,\nabla_x S) = E .  \]
Now, since $H$ is time-independent, the general solution to the Hamilton-Jacobi equation for $S$ is of the form $W(x)-Et$, and generates a canonical transformation on phase space from $(x,k)$ to a set of constants of the motion, which may be chosen as the intial values  $(x_0,k_0)$ at some fixed time \cite{goldstein}. One then has $E=H(x_t,k_t)=H(x_0,k_0)$, and a corresponding set of solutions $S_{x_0,k_0}(x,t):=W_{k_0}(x)-H(x_0,k_0)t$ of the Hamilton-Jacobi equation (where $x_0=\nabla_{k_0}W_{k_0}$ \cite{goldstein}).  Further, recalling that the classical velocity is $\dot{x}_t=\nabla_k H$, the above continuity equation simply requires that $P(x)|\dot{x}_t|$ is constant along any given trajectory $(x_t,k_t)$.  One may therefore define a corresponding set of solutions by $P_{x_0,k_0}(x)\sim |\dot{x}_t|^{-1}$ when $x$ lies on the particular trajectory having initial values $(x_0,k_0)$, and $P_{x_0,k_0}(x)=0$ elsewhere. 

The stationary ensembles  $(P_{x_0,k_0},S_{x_0,k_0})$ are all distinguishable, since they correspond to a set of distinct initial values $(x_0,k_0)$. Hence, they are suitable for defining a thermal mixture.
Further, by construction, one has via definition (\ref{cf}) that \[ C_f[P_{x_0,k_0},S_{x_0,k_0}] = \int dx_t\,|\dot{x}_t|^{-1}f(x_t,k_t){\big /}\int dx_t\,|\dot{x}_t|^{-1}, \]
where integration is along the trajectory defined by initial point $(x_0,k_0)$.  Changing the variable of integration to $t$ then gives
\begin{equation} \label{erg}
C_f[P_{x_0,k_0},S_{x_0,k_0}] = \lim_{T\rightarrow\infty} \frac{1}{T}\int_0^T dt\, f(x_t,k_t), 
\end{equation}
which is always well defined for ergodic systems.  Further, for such systems the righthand side is simply the microcanonical ensemble average of $f(x,k)$, corresponding to constant energy $E=H(x_0,k_0)$ \cite{ergodic}.  Finally, noting $\tilde{H}[P_{x_0,k_0},S_{x_0,k_0}]=H[x_0,k_0]$ by construction, the associated thermal mixture in Eq.~(\ref{mix}) is  characterised by $p(x_0,k_0)\sim e^{-\beta H[x_0,k_0]}$, and it immediately follows via Eqs.~(\ref{ava}) and (\ref{erg}) that the average of $f$ over the mixture is the usual canonical ensemble average, as required.

\section{Discussion}

The configuration-ensemble approach satisfies the two minimal requirements for a consistent formulation of mixed dynamics, as shown in Sec.~II.  No other formulation appears to be known which passes this critical test.  

Moreover, while the approach does not satisfy a strong separability condition for quantum and classical observables, it does satisfy the weaker condition of configuration separability, as per Eqs.~(\ref{cx}) and (\ref{qq}) of Sec.~III.  This condition is sufficient to avoid difficulties in the case of quantum matter coupled to a classical spacetime metric, and is also sufficient more generally if it is assumed that only the classical configuration of a mixed ensemble is directly accessible to measurement.

It should further be noted that the configuration-ensemble approach has a number of interesting applications outside the domain of mixed dynamics.  It also provides a basis for, eg, the derivation of classical and quantum equations of motion \cite{eup}, a generalisation of quantum superselection rules \cite{super}, and, as seen in Sec.~IV above, a `Hamilton-Jacobi' approach to classical statistical mechanics.  Hence, overall, the approach appears to be valuable in providing a fundamental tool for describing physical systems, and merits further general investigation.

It is of interest to remark on how the configuration-ensemble approach is able to avoid various `no-go' theorems on mixed dynamics in the literature \cite{sal1,sal2,ternoperes,sahoo}.  This is essentially due to such theorems requiring a formal assumption that the set of observables can be be extended to form a product algebra, where the product $A\ast B$ is assumed to satisfy $C_f\ast C_g=C_{fg}$, $Q_M\ast Q_N=Q_{MN}$, and some further property such as the Leibniz rule $\{A,B\ast C\} = \{A,B\}\ast C + B\ast\{A, C\}$.  However, the assumption of such a product algebra clearly goes beyond the domain of observable quantities (eg, the product of two Hermitian operators is not a Hermitian operator), and hence cannot be justified on physical grounds.  Thus, any import of such `no-go' theorems, for the configuration-ensemble approach, where no such product is defined or required, is purely formal in nature.  

For application to mixed dynamics, the set of observables must be chosen such that it contains the classical and quantum observables defined in Eqs.~(\ref{cf}) and (\ref{qm}), with all members satisfying the normalisation and positivity conditions in Eq.~(\ref{con1}).  It must also, of course, contain the Poisson bracket of any two of its members - however, this can always be assured by replacing a given set by its closure under the Poisson bracket operation.  It is of interest to consider what further physical conditions might be imposed on the set of observables.  For example, for ensembles of interacting classical and quantum nonrelativistic particles, it is reasonable to require that the equations of motion are invariant under Galilean transformations, leading to the interesting property that the centre of mass and relative motions do not decouple \cite{hr}.

A more general condition that might be imposed on observables is that they are homogenous of degree unity with respect to the probability density $P$, i.e.,
\begin{equation} \label{con2}  
A[\lambda P,S] = \lambda\,A[P,S]   
\end{equation}
for all $\lambda \geq0$.  Note that if this condition holds for two observables $A$ and $B$, then it holds for  the Poisson bracket $\{A,B\}$, as may be checked by direct substitution into Eq.~(\ref{poiss}). It is also easily verified to hold for the classical and quantum observables defined in Eqs.~(\ref{cf}) and (\ref{qm}).  Differentiating Eq.~(\ref{con2}) on both sides with respect to $\lambda$ and choosing $\lambda=1$ yields the numerical identity
\begin{equation} 
A[P,S] = \int d\xi \,P\,(\delta A/\delta P) =: \langle \delta A/\delta P\rangle , 
\end{equation}
i.e., $A$ can be calculated by integrating over a local density on the configuration space.  Thus, the homogeneity condition consistently allows observables to be interpreted both as generators of canonical transformations and as expectation values.  

Finally, while the question of decoherence has not been addressed in any detail here, it is worth noting that the configuration-ensemble approach provides at least two possibilities in this regard.  First, for any mixed quantum-classical ensemble, one may define a conditional quantum wavefunction $\psi_x(q)$ and corresponding density operator $\rho_{Q|C}$, which describe the conditional decoherence of the quantum component relative to the classical component \cite{hr}.  
Second, while at any time the hybrid wavefunction $\psi(q,x,t)$ describing a mixed quantum-classical system always has a decomposition of the form 
\[  \psi(q,x,t) = \sum_n \sqrt{p_n(t)} \, \psi_{C,n}(x,t) \, \psi_{Q,n}(q,t)  \]
(eg, a Schmidt decomposition), only at {\it particular} times (if at all, depending on the ensemble Hamiltonian), can it have such a decomposition for which (i) the classical ensembles corresponding to $\psi_{C,n}(x,t)$ are {\it classically} distinguishable (see Sec.~IV), and (ii) the quantum ensembles corresponding to $\psi_{Q,n}(q,t)$ are mutually orthogonal.  Moreover, unlike a Schmidt decomposition, such a decomposition would be unique even for equal $p_n(t)$. Hence, decoherence could be modelled by imposing a spontaneous `collapse' of the ensemble at such well-defined times, similarly to the collapse model in Sec.~III~C.

\begin{acknowledgments}
I am grateful to Marcel Reginatto for valuable discussions, and for pointing out an error in an early draft.
\end{acknowledgments}

\end{document}